\documentclass{article}
\usepackage{epsfig}
\usepackage{amssymb}
\usepackage{amsmath}
\usepackage{color}
\usepackage{subfigure}
\newtheorem{rem}{Remark}[section]
\newcommand{\br}{\begin{rem}}
\newcommand{\er}{\end{rem}}
\newtheorem{ex}[rem]{Example}
\newcommand{\bex}{\begin{ex}}
\newcommand{\eex}{\end{ex}}
\newtheorem{Def}[rem]{Definition}
\newcommand{\bd}{\begin{Def}}
\newcommand{\ed}{\end{Def}}
\newtheorem{theorem}[rem]{Theorem}
\newcommand{\bt}{\begin{theorem}}
\newcommand{\et}{\end{theorem}}
\newtheorem{prop}[rem]{Proposition}
\newcommand{\bp}{\begin{prop}}
\newcommand{\ep}{\end{prop}}
\newtheorem{lemma}[rem]{Lemma}
\newcommand{\bl}{\begin{lemma}}
\newcommand{\el}{\end{lemma}}
\newtheorem{corollary}[rem]{Corollary}
\newcommand{\bc}{\begin{corollary}}
\newcommand{\ec}{\end{corollary}}
\newcommand{\be}{\begin{equation}}
\newcommand{\ee}{\end{equation}}
\newcommand{\bea}{\begin{eqnarray}}
\newcommand{\eea}{\end{eqnarray}}
\newcommand{\pa}{\partial}
\newcommand{\nn}{\nonumber}
\newcommand{\adots}{\mathinner{\mkern2mu\raise1pt\hbox{.}\mkern2mu
\raise4pt\hbox{.}\mkern2mu\raise7pt\hbox{.}\mkern1mu}}

\title{Integrable Maps which Preserve Functions with Symmetries}
\author{Allan P Fordy\thanks{School of Mathematics,
University of Leeds, Leeds LS2 9JT. ~~E-mail: A.P.Fordy@leeds.ac.uk}
$\,$ and Pavlos Kassotakis\thanks{School of Mathematics and Statistics,
 Carslaw Building (F07), University of Sydney, NSW 2006, Australia.
 E-mail: P.Kassotakis@maths.usyd.edu.au, pavlos78@gmail.com}}

\begin{document}

\maketitle

\begin{abstract}
We consider maps which preserve functions which are built out of the invariants of some simple vector fields.
We give a reduction procedure, which can be used to derive commuting maps of the plane, which preserve the {\em same} symplectic form and first integral. We show how our method can be applied to some maps which have recently appeared in the context of Yang-Baxter maps.
\end{abstract}


\section{Introduction}

In this paper we consider a class of mappings on a $2n-$dimensional space, which preserve $n$ functions which are built out of the invariants of some simple vector fields (which are therefore {\em symmetries} of the ``first integrals'' of the mappings).  We describe circumstances in which the $2n-$dimensional map reduces to a symplectic map of the plane, possessing a first integral.  Our examples are for $n=2$ and $n=3$ and the invariant functions on the plane are of QRT type \cite{88-5}.  However, depending upon the choice of the ``simple vector field'', the induced map of the plane is {\em different}, whilst the invariant function and symplectic form are the {\em same}.  We thus construct pairs of maps of the plane which preserve the {\em same} function.  We find that these maps commute.  We can also build the QRT map corresponding to the reduced integral and this also commutes with our maps.  In fact, in most of our examples, one of our reduced maps {\em coincides} with this QRT map, but Example \ref{mcm4d-2} shows that this need not be the case.

We first explain the method within the context of a $4-$dimensional generalised McMillan map introduced in \cite{f06-2}.  After this example we discuss some general ideas and features in Section \ref{general4d}, and illustrate these with two more examples of McMillan type.  We then consider two maps which recently appeared \cite{09-4,12-1} in the context of Yang-Baxter maps, one of which is the Adler-Yamilov map \cite{94-7}.

In the $6-$dimensional context, we need $3$ functions and an additional symmetry vector field.  In Section \ref{6d}, we give a generalisation of one of our $4-$dimensional examples to show how this can be done.

Whilst our current examples are in $4$ and $6$ dimensions, the method can, in principle, be used to reduce a class of $2n-$dimensional maps with $n$ invariant functions and $n-1$ symmetries.
This should be compared with the reduction method of \cite{99-10}, where $2n-1$ invariant functions are required, $2n-2$ of which are Casimirs of the resulting Poisson bracket, which then reduces to $2-$dimensions.

In Section \ref{summary} we discuss some general features of the method, as well as some of the difficulties that might arise in higher dimensions.

\section{The $4$ Dimensional Case}

We describe the basic method within the context of the next example.

\bex[A $4d$ McMillan Map \cite{f06-2}] \label{mcm4d} {\em  %
Consider the vector field ${\bf X} = (x_1,-x_2,-y_1,y_2)$, which has invariants
\be\label{gammai1}  %
\gamma_1=x_1y_1,\;\; \gamma_2=x_2y_2,\;\; \gamma_3=x_1x_2,\;\; \gamma_4=y_1y_2,\quad\mbox{satisfying}\;\; \gamma_3\gamma_4=\gamma_1\gamma_2.
\ee  %
Under the involution
\be  \label{iotaxy}  %
\iota_{xy}: (x_i,y_i)\mapsto (y_i,x_i), \quad i=1,2,
\ee  %
we have $\gamma_3\leftrightarrow \gamma_4$ and $\gamma_1, \gamma_2$ invariant, so the following functions are invariant under this:
\be  \label{h1h2}  %
h_1 = (1-\gamma_3)(1-\gamma_4)-2a\gamma_1=(1-x_1 x_2)(1-y_1y_2)-2 a x_1 y_1, \quad h_2=\gamma_1-\gamma_2=x_1y_1-x_2y_2.
\ee  %
The equations
$$
h_k(\tilde {\bf x},{\bf y})=h_k({\bf x},{\bf y}),\quad k=1,2,
$$
can be solved for $\tilde x_i,\; i=1,2$, to obtain the involution $\rho_{x}$ and this can be composed with $\iota_{xy}$ to give
$$
\varphi=\iota_{xy}\circ\rho_{x}:({\bf x},{\bf y})\mapsto
\left(y_1,y_2,-\frac{x_2y_2}{y_1}-\frac{2ay_2}{1-y_1y_2}, -\frac{x_1y_1}{y_2}-\frac{2ay_1}{1-y_1y_2}\right),
$$
which is a coupled McMillan map.  The Jacobian of this map is $-1$, so the volume form $\Omega_4=dx_1\wedge dx_2\wedge dy_1 \wedge dy_2$ is {\em anti-invariant}.  Under the map, the vector ${\bf X}\mapsto -{\bf X}$, so the $3-$form $\Omega_3={\bf X}\;\lrcorner\; \Omega_4$ is {\em invariant}.

At this stage we change coordinates:
$$
u_1=\gamma_3,\; v_1=\gamma_1,\; u_2=h_2,\; v_2=y_2, \quad\mbox{with Jacobian}\;\; x_1x_2y_2=u_1v_2.
$$
In these coordinates ${\bf X}=v_2\pa_{v_2}$, so
$$
\Omega_3={\bf X}\;\lrcorner\; \Omega_4=-\frac{du_1\wedge dv_1 \wedge du_2}{u_1}.
$$
It is important for us that this form is {\em closed}.  The map $\varphi$ now takes the form
\bea  %
&& \tilde u_1 = \frac{v_1(v_1-u_2)}{u_1}, \quad \tilde v_1= \frac{(u_2-v_1)(u_1+(2a+u_2-v_1)v_1)}{u_1+(u_2-v_1) v_1},\nn\\
\label{4duvmap}  \\
&& \tilde u_2=u_2,   \quad  \tilde v_2=\frac{((u_2-v_1)(2a-v_1)-u_1)v_1}{(u_1+(u_2-v_1) v_1)v_2}.  \nn
\eea  %
On the level surface $u_2=k$, the map restricts to the $u_1, v_1$ plane,
\be\label{utilde1}  %
\tilde u_1 = \frac{v_1(v_1-k)}{u_1},\;\; \tilde v_1= \frac{(k-v_1)(u_1+(2a+k-v_1)v_1)}{u_1+(k-v_1) v_1},
\ee  %
with {\em invariant function} $h_1$ and {\em invariant symplectic form} $\omega$, given by
\be\label{huwu}  %
h_1=1-u_1-(2a+k)v_1+v_1^2+\frac{(k-v_1)v_1}{u_1}, \quad \omega = \frac{du_1\wedge dv_1}{u_1}.
\ee  %
This two dimensional map is just the QRT map built from the two involutions $h_1(\tilde u_1,v_1)=h_1(u_1,v_1)$ and $h_1(u_1,\tilde v_1)=h_1(u_1,v_1)$.

\medskip
Now consider ${\bf X} = (x_1,x_2,-y_1,-y_2)$, with invariants
\be\label{gammai2}  %
\gamma_1=x_1y_1,\;\; \gamma_2=x_2y_2,\;\; \gamma_3=x_1y_2,\;\; \gamma_4=x_2y_1,\quad\mbox{also satisfying}\;\; \gamma_3\gamma_4=\gamma_1\gamma_2,
\ee  %
with $\gamma_3\leftrightarrow \gamma_4$ and $\gamma_1, \gamma_2$ invariant under the action of $\iota_{xy}$.  Taking the {\em same} functions of $\gamma_i$,
$$
h_1 = (1-\gamma_3)(1-\gamma_4)-2a\gamma_1=(1-x_1 y_2)(1-x_2y_1)-2 a x_1 y_1, \quad h_2=\gamma_1-\gamma_2=x_1y_1-x_2y_2,
$$
and proceeding as before, we find
$$
\hat\varphi:({\bf x},{\bf y})\mapsto
\left(y_1,y_2,\frac{2a}{y_1}+\frac{1}{y_2}+\frac{y_2(1-x_2y_1)}{y_1^2}, \frac{2a}{y_2}+\frac{1}{y_1}+\frac{y_1(1-x_1y_2)}{y_2^2}\right),
$$
which again has the {\em anti-invariant} volume form is $\Omega_4=dx_1\wedge dx_2\wedge dy_1 \wedge dy_2$, and gives ${\bf X}\mapsto -{\bf X}$.

Again, using the change of coordinates:
$$
u_1=\gamma_3=x_1 y_2,\; v_1=\gamma_1=x_1y_1,\; u_2=h_2=x_1y_1-x_2y_2,\; v_2=y_2,
$$
we find a reduced $2-$dimensional map (on the level surface $u_2=k$):
\be\label{uhat1}  %
\hat u_1 =  \frac{u_1^2+v_1^2+u_1v_1(2a-v_1)}{u_1^2},\;\; \hat v_1= 2a+k-v_1+\frac{u_1}{v_1}+\frac{v_1}{u_1},
\ee  %
which again preserves $h_1$ and $\omega$ of (\ref{huwu}).  On this $2-$dimensional space, $\varphi$ and $\hat\varphi$ commute.
}\eex

\subsection{Some General Theory}\label{general4d}

With this example in mind, we can develop some general theory (for the moment, we take $n=2$).  We look more closely at some of the basic ingredients of our construction.

\paragraph{The vector field $\bf X$ and its invariants:} The pair of functions $h_1, h_2$ are built out of these invariants.
Before even constructing the map $\varphi$ our choice of invariants will dictate the form of the reduced function $h_1$.

For instance, starting with ${\bf X}=(x_1,-x_2,-y_1,y_2)$, we can {\em initially} choose $\gamma_1=x_1y_1,\, \gamma_2=x_2y_2,\, \gamma_3=x_1x_2,\, \gamma_4=y_1y_2$, which satisfy $\gamma_3\gamma_4=\gamma_1\gamma_2$.  Our construction also uses a simple involution, under which ${\bf X}\mapsto \pm {\bf X}$.  In this paper we use one of the following involutions:
$$
\iota_{xy}: (x_i,y_i)\mapsto (y_i,x_i), \;\; i=1,2,\quad\mbox{or}\quad \iota_{12}: (x_1,x_2,y_1,y_2)\mapsto (x_2,x_1,y_2,y_1).
$$
We have
\be\label{iotagamma}  %
\iota_{xy}: (\gamma_1,\gamma_2,\gamma_3,\gamma_4)\mapsto (\gamma_1,\gamma_2,\gamma_4,\gamma_3) \quad\mbox{and}\quad
   \iota_{12}: (\gamma_1,\gamma_2,\gamma_3,\gamma_4)\mapsto (\gamma_2,\gamma_1,\gamma_3,\gamma_4).
\ee  %
In Example \ref{mcm4d} we used $\iota_{xy}$ and wrote two functions $h_1,h_2$, given by (\ref{h1h2}), which were invariant under this involution.
The constraint $\gamma_3\gamma_4=\gamma_1\gamma_2$, together with the reduction to a level surface $h_2=k$, then gives us
$$
h_1=(1-\gamma_3)\left(1-\frac{\gamma_1(\gamma_1-k)}{\gamma_3}\right)-2a\gamma_1,
$$
written in terms of two variables $\gamma_1, \gamma_3$.

Our choice of $\gamma_i$ was rather arbitrary.  Clearly
$$
\hat\gamma_1=\gamma_1,\;\; \hat\gamma_2=\gamma_2,\;\; \hat\gamma_3=\frac{\gamma_3}{\gamma_1},\;\; \hat\gamma_4=\frac{\gamma_4}{\gamma_1}
$$
also satisfy
$\iota_{xy}: (\hat\gamma_1,\hat\gamma_2,\hat\gamma_3,\hat\gamma_4)\mapsto (\hat\gamma_1,\hat\gamma_2,\hat\gamma_4,\hat\gamma_3)$,
so we could use these instead.  The same description of $h_1, h_2$ in terms of $\gamma_i$ can be used for either choice (to be invariant under $\iota_{xy}$), but now we have a {\em different} constraint, $\hat\gamma_3\hat\gamma_4=\hat\gamma_2/\hat\gamma_1$, which leads to a {\em different} reduction to two dimensions.  This will be illustrated in Example \ref{mcm4d-2} below.

For each example we can calculate $\rho_{x}$, as we did in Example \ref{mcm4d} and then define $\varphi=\iota_{xy}\circ\rho_{x}$.
Any map which preserves the functions $h_i$ will map (through its Jacobian) a symmetry vector of $h_i$ to a symmetry vector.
For the case $\rho_{x}:(x_1,x_2,y_1,y_2)\mapsto (\tilde x_1,\tilde x_2,y_1,y_2)$, we have that the third and fourth components of our vector $\bf X$ remain invariant, so ${\bf X}\mapsto {\bf X}$.  This gives us the important property that under $\varphi$, ${\bf X}\mapsto \pm {\bf X}$.

\br[Choice of $h_i$]  %
The choice of functions $h_1, h_2$ is initially only constrained by the invariance under the simple involution.  However, the forms of $h_i$ are further restricted by the requirement that $\rho_{x}$ is rational.
\er  %

\paragraph{Measure preserving maps:}

Another ingredient is an invariant (or anti-invariant) volume form $\Omega_4$, which leads to the invariant form $\Omega_3={\bf X}\;\lrcorner\; \Omega_4$ and to the invariant symplectic form on the $2-$dimensional reduction.

In order to guarantee that $d\Omega_3=0$, we need that the Lie derivative $L_X\Omega_4=0$, since
$$
L_X\Omega_4=d({\bf X}\;\lrcorner\; \Omega_4)+{\bf X}\;\lrcorner\; d\Omega_4=d({\bf X}\;\lrcorner\; \Omega_4).
$$
This gives a divergence condition on $\bf X$, which is the $n=2$ case of the following:
\be\label{divX}  %
\Omega=\frac{dx_1\wedge\cdots\wedge dy_n}{\sigma (x_1,\dots ,y_n)}\quad\Rightarrow\quad
d({\bf X}\;\lrcorner\; \Omega)=\sum_{i=1}^{n} \frac{\pa }{\pa x_i}\left(\frac{{\bf X}^i}{\sigma}\right)+\sum_{i=1}^{n} \frac{\pa }{\pa y_i}\left(\frac{{\bf X}^{n+i}}{\sigma}\right)=0.
\ee  %
All our initial vector fields satisfy $\sum_{i=1}^n(\pa_{x_i} X_i+\pa_{y_i} X_{n+i})=0$,
which means that ${\bf X}\cdot \nabla (\sigma)=0$, so $\sigma$ must also be built out of the invariants of $\bf X$.  This means that ${\bf X}\;\lrcorner\; \Omega$ can be written in terms of the invariants of $\bf X$.

\paragraph{Reduction to the space of invariants:}

Since the vector ${\bf X}\mapsto \pm {\bf X}$ under the map $\varphi$, the space of invariants $\gamma_i$, must itself be invariant under the action of $\varphi$.  This is why the map (\ref{4duvmap}) splits into a $3-$dimensional map on its space of invariants, together with a map of $v_2$, with coefficients depending on the other variables.  Furthermore, since $v_2$ is chosen to make ${\bf X}=\pm v_2\pa_{v_2}$, we have
$$
v_2\pa_{v_2}=v_2\frac{\pa \tilde v_2}{\pa v_2} \frac{\pa}{\pa \tilde v_2},
$$
and since ${\bf X}\mapsto \pm {\bf X}$, we have $v_2\pa_{v_2} \tilde v_2=\pm \tilde v_2$ and hence $\tilde v_2=\alpha v_2^{\pm 1}$, where $\alpha$ is a function of $u_1,v_1,u_2$.  The $3-$dimensional map on the space of invariants again decouples into a $2-$dimensional map, together with the identity map on the third component.

\paragraph{The second symmetry:}  for any pair of functions $h_1,\, h_2$ we can construct a pair of symmetry vectors $\sigma_1,\, \sigma_2$ (the $4d$ equivalent of (\ref{syms})), a linear combination of which gives us back our original vector $\bf X$.  The Hamiltonian vector field generated by the reduced $h_1$ is also a symmetry vector, so can itself be written in terms of our basis elements $\sigma_i$.

\paragraph{Higher dimensions:}

In higher dimensions we require the same ingredients, together with additional vector fields which satisfy ${\bf X}_i h_j=0$, each of which should satisfy ${\bf X}_i\mapsto \pm {\bf X}_i$, together with the above divergence condition.

\section{Some Further Examples in $4D$}

We give brief details of two more McMillan type maps.  The first uses the {\em same} form of $h_i$, but changes the form of $\gamma_i$.  Example \ref{mcm4d-3} changes the form of $h_2$ but uses our standard set of $\gamma_i$.

\bex[The $4d$ Map of Example (\ref{mcm4d}) with another choice of $\gamma_i$] \label{mcm4d-2} {\em  %
Consider again the vector field ${\bf X} = (x_1,-x_2,-y_1,y_2)$.  We replace the invariants (\ref{gammai1}) by
$$
\hat\gamma_1=\gamma_1,\;\; \hat\gamma_2=\gamma_2,\;\; \hat\gamma_3=\frac{\gamma_3}{\gamma_1},\;\; \hat\gamma_4=\frac{\gamma_4}{\gamma_1}
$$
and then ``drop the hats''.  We thus have
$$
\gamma_1=x_1y_1,\;\; \gamma_2=x_2y_2,\;\; \gamma_3=\frac{x_2}{y_1},\;\; \gamma_4=\frac{y_2}{x_1},\quad\mbox{satisfying}\;\; \gamma_3\gamma_4=\frac{\gamma_2}{\gamma_1}.
$$
Under the involution $\iota_{xy}$, we still have $\gamma_3\leftrightarrow \gamma_4$ and $\gamma_1, \gamma_2$ invariant, so the same functions $h_i$, given by (\ref{h1h2}) are invariant under this.  However, they now take the explicit form
\be  \label{h1h2-2}  %
h_1 = \left(1-\frac{x_2}{y_1}\right)\left(1-\frac{y_2}{x_1}\right)-2 a x_1 y_1, \quad h_2=x_1y_1-x_2y_2.
\ee  %
We may follow the same procedure and again change to coordinates:
$$
u_1=\gamma_3,\; v_1=\gamma_1,\; u_2=h_2,\; v_2=y_2,
$$
which leads us (on the level surface $u_2=k$) to the map $\varphi_1$, given by
\be\label{utilde1-2}  %
\tilde u_1 = \frac{u_1v_1}{v_1+k (u_1-1)}+\frac{2a v_1(v_1-k)}{v_1+k (u_1-1)},\;\; \tilde v_1= \frac{(v_1-k)(v_1+k (u_1-1))}{u_1v_1(u_1+2a(v_1-k))}.
\ee  %
The constraint $\gamma_3\gamma_4=\gamma_2/\gamma_1$ leads to a {\em different} reduction of $h_1$:
\be\label{huwu-2}  %
h_1=\frac{k(1-u_1)}{u_1v_1}-\frac{(1-u_1)^2}{u_1}-2av_1.
\ee  %
We find that this map leaves invariant the symplectic form
$$
\omega = \frac{du_1\wedge dv_1}{u_1v_1}.
$$
The QRT map built from the two involutions $h_1(\tilde u_1,v_1)=h_1(u_1,v_1)$ and $h_1(u_1,\tilde v_1)=h_1(u_1,v_1)$ is now
\be\label{qrt1}  %
Q:(u_1,v_1)\mapsto \left(\frac{v_1-k}{u_1v_1},\frac{k(k+v_1(u_1-1))}{2av_1(k-v_1)}\right),
\ee  %
which is different from $\varphi_1$.  These maps commute.

\medskip
We now repeat the process using ${\bf X} = (x_1,x_2,-y_1,-y_2)$, and define $\hat \gamma_i$ in the same way, but with $\gamma_i$ given by (\ref{gammai2}).  Dropping hats, we find
$$
\gamma_1=x_1y_1,\;\; \gamma_2=x_2y_2,\;\; \gamma_3=\frac{y_2}{y_1},\;\; \gamma_4=\frac{x_2}{x_1},\quad\mbox{satisfying}\;\; \gamma_3\gamma_4=\frac{\gamma_2}{\gamma_1}.
$$
The function $h_i$ now take the explicit form
\be  \label{h1h2-22}  %
h_1 = \left(1-\frac{y_2}{y_1}\right)\left(1-\frac{x_2}{x_1}\right)-2 a x_1 y_1, \quad h_2=x_1y_1-x_2y_2.
\ee  %
Again, using coordinates
$$
u_1=\gamma_3,\; v_1=\gamma_1,\; u_2=h_2,\; v_2=y_2,
$$
leads us (on the level surface $u_2=k$) to the map $\varphi_2$, given by
\be\label{utilde1-22}  %
\bar u_1 = \frac{1}{u_1}-\frac{2av_1}{u_1-1},\;\; \bar v_1= \frac{k(u_1-1)}{2au_1v_1},
\ee  %
which preserves the same reduced $h_1$ and the symplectic form $\omega$.

On this $2-$dimensional space, $\varphi_1,\, \varphi_2$ and $Q$ pairwise commute.
}\eex

\bex[A $4d$ McMillan Map with another choice of $h_2$] \label{mcm4d-3} {\em  %

Starting again with the vector field ${\bf X} = (x_1,-x_2,-y_1,y_2)$, with invariants (\ref{gammai1}), we take the two functions
$$
h_1 = (1-\gamma_3)(1-\gamma_4)-2a\gamma_1=(1-x_1 x_2)(1-y_1y_2)-2 a x_1 y_1, \quad h_2=\frac{\gamma_1}{\gamma_2}=\frac{x_1y_1}{x_2y_2},
$$
which are invariant under the involution $\iota_{xy}$.

We may follow the same procedure and again change to coordinates:
$$
u_1=\gamma_3,\; v_1=\gamma_1,\; u_2=h_2,\; v_2=y_2,
$$
which leads us (on the level surface $u_2=k$) to the map $\varphi_1$, given by
$$
\tilde u_1 = \frac{v_1^2}{ku_1},\;\; \tilde v_1= -v_1-\frac{2akv_1^2}{ku_1-v_1^2},
$$
with {\em invariant function} $h_1$ and {\em invariant symplectic form} $\omega$, given by
\be\label{huwu2}  %
h_1=1-u_1-2av_1+\frac{(u_1-1)v_1^2}{ku_1}, \quad \omega = \frac{du_1\wedge dv_1}{u_1}.
\ee  %
This two dimensional map is just the QRT map built from the two involutions $h_1(\tilde u_1,v_1)=h_1(u_1,v_1)$ and $h_1(u_1,\tilde v_1)=h_1(u_1,v_1)$.

Now consider ${\bf X} = (x_1,x_2,-y_1,-y_2)$, with invariants (\ref{gammai2}).  Taking the {\em same} functions of $\gamma_i$,
$$
h_1 = (1-\gamma_3)(1-\gamma_4)-2a\gamma_1=(1-x_1 y_2)(1-x_2y_1)-2 a x_1 y_1, \quad h_2=\frac{\gamma_1}{\gamma_2}=\frac{x_1y_1}{x_2y_2},
$$
we may follow the same procedure and again change to coordinates:
$$
u_1=\gamma_3,\; v_1=\gamma_1,\; u_2=h_2,\; v_2=y_2,
$$
which leads us (on the level surface $u_2=k$) to the map $\varphi_2$, given by
\be\label{uhat2}  %
\hat u_1 = 1+\frac{(1-u_1)v_1^2}{ku_1^2}+\frac{2av_1}{u_1},\;\; \hat v_1= -v_1+\frac{ku_1}{v_1}+\frac{v_1}{u_1}+2ak,
\ee  %
which again preserves $h_1$ and $\omega$ of (\ref{huwu2}).  On this $2-$dimensional space, $\varphi_1$ and $\varphi_2$ commute.
}\eex

\subsection{Reductions of some Yang-Baxter Maps}

In this section we take two Yang-Baxter maps which have appeared in the literature and show that they can be considered in the framework of this paper.

\bex[The Adler-Yamilov Map and a Modification]\label{ay-ex}  {\em  %
Starting with
$$
{\bf X} = (x_1,-x_2,y_1,-y_2),\quad\mbox{with invariants}\quad
   \gamma_1=x_1x_2,\;\; \gamma_2=y_1y_2,\;\; \gamma_3=x_1y_2,\;\; \gamma_4=x_2y_1,
$$
satisfying $\gamma_3\gamma_4=\gamma_1\gamma_2$, we may consider any pair of functions built out of these invariants.  To arrive at the Adler-Yamilov map \cite{94-7,09-4}, we use the invariant functions given in \cite{12-1}:
\bea  %
h_1 &=& \gamma_1+\gamma_2= x_1 x_2+y_1y_2, \nn\\
 h_2 &=& a\gamma_2+b\gamma_1+\gamma_3+\gamma_4+\gamma_1\gamma_2= ay_1y_2+bx_1x_2+x_1y_2+x_2y_1+x_1x_2y_1y_2. \nn
\eea  %
The functions $h_k$ are clearly invariant under the action of $\iota_{12}: (x_1,x_2,y_1,y_2)\mapsto (x_2,x_1,y_2,y_1)$.  The equations $h_k(\tilde x_1,y_1,x_2,\tilde y_2)=h_k(x_1,x_2,y_1,y_2)$ give us the involution
$$
\rho_{x_1y_2}:  (x_1,x_2,y_1,y_2)\mapsto \left(y_2-\frac{(a-b)x_2}{1+x_2y_1},y_1,x_2,x_1+\frac{(a-b)y_1}{1+x_2y_1}\right).
$$
The Adler-Yamilov map is obtained from their composition:
$$
\varphi=\rho_{x_1y_2}\circ \iota_{12}:(x_1,x_2,y_1,y_2)\mapsto \left(y_1-\frac{(a-b)x_1}{1+x_1y_2},y_2,x_1,x_2+\frac{(a-b)y_2}{1+x_1y_2}\right).
$$
The volume form $\Omega_4=dx_1\wedge dx_2\wedge dy_1 \wedge dy_2$ is invariant under this map and the vector ${\bf X}\mapsto {\bf X}$, so the $3-$form $\Omega_3={\bf X}\;\lrcorner\; \Omega_4$ is invariant.

At this stage we change coordinates:
$$
u_1=\gamma_1,\; v_1=\gamma_3,\; u_2=h_1,\; v_2=y_2, \quad\mbox{with Jacobian}\;\; -x_1y_2^2=-v_1v_2.
$$
In these coordinates ${\bf X}=-v_2\pa_{v_2}$, so
\be\label{ay-Omega3}  %
\Omega_3={\bf X}\;\lrcorner\; \Omega_4=-\frac{du_1\wedge dv_1 \wedge du_2}{v_1},
\ee   %
which is evidently closed.  The map $\varphi$ now takes the form
\bea  %
&&  \tilde u_1 = u_2-u_1-\frac{(a-b)v_1}{1+v_1},\;\;
    \tilde v_1= \frac{1}{v_1}\left(u_1+\frac{(a-b)v_1}{1+v_1}\right)\left(u_1-u_2+\frac{(a-b)v_1}{1+v_1}\right),\nn\\
    &&   \label{aymap-uv}  \\
&&   \tilde u_2=u_2,\;\; \tilde v_2=\frac{v_2}{v_1}\left(u_1+\frac{(a-b)v_1}{1+v_1}\right).  \nn
\eea  %
On the level surface $u_2=k$, the map restricts to the coordinates $u_1, v_1$:
$$
\tilde u_1 = k-u_1-\frac{(a-b)v_1}{1+v_1},\;\;
   \tilde v_1= \frac{1}{v_1}\left(u_1+\frac{(a-b)v_1}{1+v_1}\right)\left(u_1-k+\frac{(a-b)v_1}{1+v_1}\right),
$$
with {\em invariant function} $h_2$ and {\em invariant symplectic form} $\omega$, given by
\be\label{ay-omega}  %
h_2= ak+v_1+\frac{u_1(k-u_1)(1+v_1)}{v_1}-(a-b)u_1,\quad \omega = \frac{du_1\wedge dv_1}{v_1}.
\ee  %
This two dimensional map is just the QRT map built from the two involutions $h_2(\tilde u_1,v_1)=h_2(u_1,v_1)$ and $h_2(u_1,\tilde v_1)=h_2(u_1,v_1)$.

It is known \cite{09-4} that the Adler-Yamilov map is symplectic, with $\omega= dx_1\wedge dx_2 + dy_1\wedge dy_2$, which, when written in terms of coordinates $u_i, v_i$, is just
$$
\omega = \frac{du_2\wedge dv_2}{v_2}-\frac{du_1\wedge dv_1}{v_1},
$$
which is invariant under the map (\ref{aymap-uv}).  The above two dimensional symplectic form is just the restriction of this to the $u_1,v_1$ space.

We may now consider
$$
{\bf X} = (x_1,-x_2,-y_1,y_2),\quad\mbox{with invariants}\quad
   \gamma_1=x_1x_2,\;\; \gamma_2=y_1y_2,\;\; \gamma_3=x_1y_1,\;\; \gamma_4=x_2y_2,
$$
satisfying $\gamma_3\gamma_4=\gamma_1\gamma_2$.  The functions $h_1, h_2$ now take the form
\bea  %
h_1 &=& \gamma_1+\gamma_2= x_1 x_2+y_1y_2, \nn\\
 h_2 &=& a\gamma_2+b\gamma_1+\gamma_3+\gamma_4+\gamma_1\gamma_2= ay_1y_2+bx_1x_2+x_1y_1+x_2y_2+x_1x_2y_1y_2. \nn
\eea  %
If we define $\hat\rho_{x_1y_2}$ as a solution of $h_k(\hat x_1,x_2,y_1,\hat y_2)=h_k(x_1,x_2,y_1,y_2)$, then we have
$$
\hat \varphi=\hat\rho_{x_1y_2}\circ \iota_{12}:(x_1,x_2,y_1,y_2)\mapsto
\left(\frac{(1+x_1y_1)y_2}{x_1^2}-\frac{1}{y_2}-\frac{(a-b)}{x_1},x_1,y_2,\frac{x_1x_2}{y_2}+\frac{x_1}{y_2^2}-\frac{1}{x_1}+\frac{(a-b)}{y_2}\right).
$$
The volume form
$
\Omega_4=dx_1\wedge dx_2\wedge dy_1 \wedge dy_2
$
is now {\em anti}-invariant, but the vector ${\bf X}\mapsto -{\bf X}$, so the $3-$form $\Omega_3={\bf X}\;\lrcorner\; \Omega_4$ is invariant.
In the coordinates:
$$
u_1=\gamma_1,\; v_1=\gamma_3,\; u_2=h_1,\; v_2=y_2, \quad\mbox{with Jacobian}\;\; -x_1y_1y_2=-v_1v_2,
$$
we have ${\bf X}=v_2\pa_{v_2}$, so we find the same formula (\ref{ay-Omega3}) for $\Omega_3$.

On the level surface $u_2=k$, the map $\hat\varphi$ restricts to the coordinates $u_1, v_1$:
$$
\hat u_1 = b-a+k-u_1+\frac{k-u_1}{v_1}-\frac{v_1}{k-u_1},\;\;
   \hat v_1= (k-u_1)^2\left(\frac{v_1+1}{v_1^2}\right)-\frac{(a-b)(k-u_1)}{v_1}-1,
$$
with the same {\em invariant function} $h_2$ and {\em invariant symplectic form} $\omega$, given by (\ref{ay-omega}).
On this $2-$dimensional space, the maps $\varphi$ and $\hat\varphi$ commute.
}\eex

\bex[Another Yang-Baxter Map]\label{yb2-ex}  {\em  %
Here we wish to consider the map (38), presented in \cite{12-1}, so again start with
$$
{\bf X} = (x_1,-x_2,y_1,-y_2),\quad\mbox{with invariants}\quad
   \gamma_1=x_1x_2,\;\; \gamma_2=y_1y_2,\;\; \gamma_3=x_1y_2,\;\; \gamma_4=x_2y_1,
$$
satisfying $\gamma_3\gamma_4=\gamma_1\gamma_2$.  We use the invariant functions given in \cite{12-1}:
\bea  %
h_1 &=& \gamma_1+\gamma_2+\gamma_3+\gamma_4= x_1 x_2+y_1y_2+x_1y_2+x_2y_1, \nn\\
 h_2 &=& a\gamma_2+b\gamma_1+\gamma_1\gamma_2= ay_1y_2+bx_1x_2+x_1x_2y_1y_2, \nn
\eea  %
which are again invariant under the action of $\iota_{12}: (x_1,x_2,y_1,y_2)\mapsto (x_2,x_1,y_2,y_1)$.  The equations $h_i(\kappa y_2,\tilde x_2, \tilde y_1,x_1/\kappa)=h_i(x_1,x_2,y_1,y_2)$ (with $\kappa$ a constant) give us
$$
\tilde x_2=\left(\frac{a-x_1y_2}{b-x_1y_2}\right)\frac{y_1}{\kappa}+\frac{(a-b)x_1}{(b-x_1y_2)\kappa}, \quad
\tilde y_1=\left(\frac{b-x_1y_2}{a-x_1y_2}\right)\kappa x_2-\frac{(a-b)\kappa y_2}{a-x_1y_2}.
$$
Noting that $x_1y_2$ is invariant under this involution, we may choose $\kappa=\frac{a-x_1y_2}{b-x_1y_2}$.  This defines the involution $\rho_{x_2y_1}$.  The composition $\varphi=\iota_{12}\circ \rho_{x_2y_1}$ gives us map (38) of \cite{12-1}:
$$
\varphi:(x_1,x_2,y_1,y_2)\mapsto \left(y_1+\frac{(a-b)x_1}{a-x_1y_2},\frac{(a-x_1y_2)y_2}{b-x_1y_2}, \frac{(b-x_1y_2) x_1}{a-x_1y_2}, x_2-\frac{(a-b)y_2}{b-x_1y_2}\right).
$$
The volume form $\Omega_4=dx_1\wedge dx_2\wedge dy_1 \wedge dy_2$ is invariant under the map and the vector ${\bf X}\mapsto {\bf X}$, so the $3-$form $\Omega_3={\bf X}\;\lrcorner\; \Omega_4$ is invariant.

The change of coordinates:
$$
u_1=\gamma_1,\; v_1=\gamma_3,\; u_2=h_1,\; v_2=y_2, \quad\mbox{with Jacobian}\;\; -x_1y_2(x_2+y_2)=-(u_1+v_1)v_2,
$$
and ${\bf X}=-v_2\pa_{v_2}$, gives
$$
\Omega_3={\bf X}\;\lrcorner\; \Omega_4=-\frac{du_1\wedge dv_1 \wedge du_2}{u_1+v_1},
$$
which is evidently closed.  On the level surface $u_2=k$, the map restricts to the coordinates $u_1, v_1$:
\bea  %
\tilde u_1 &=& k-v_1-\frac{(a+u_1)ku_1}{(b+u_1)(u_1+v_1)}+\frac{(a-b)bk}{(b+u_1)(b-v_1)},\nn\\
   \tilde v_1 &=& b-a-u_1+\frac{ku_1(a+u_1)}{(b+u_1)(u_1+v_1)}+(a-b)\left(\frac{a+u_1}{a-v_1}-\frac{bk}{(b+u_1)(b-v_1)}\right),\nn
\eea  %
with {\em invariant function} $h_2$ and {\em invariant symplectic form} $\omega$, given by
$$
h_2=a(k-v_1)+(b+k)u_1-u_1v_1-\frac{(a+u_1)ku_1}{u_1+v_1},\quad \omega = \frac{du_1\wedge dv_1}{u_1+v_1}.
$$
This two dimensional map is just the QRT map built from the two involutions $h_2(\tilde u_1,v_1)=h_2(u_1,v_1)$ and $h_2(u_1,\tilde v_1)=h_2(u_1,v_1)$.

We may now consider
$$
{\bf X} = (x_1,-x_2,-y_1,y_2),\quad\mbox{with invariants}\quad
   \gamma_1=x_1x_2,\;\; \gamma_2=y_1y_2,\;\; \gamma_3=x_1y_1,\;\; \gamma_4=x_2y_2,
$$
satisfying $\gamma_3\gamma_4=\gamma_1\gamma_2$.  The functions $h_1, h_2$ now take the form
\bea  %
h_1 &=& \gamma_1+\gamma_2+\gamma_3+\gamma_4= x_1 x_2+y_1y_2+x_1y_1+x_2y_2, \nn\\
 h_2 &=& a\gamma_2+b\gamma_1+\gamma_1\gamma_2= ay_1y_2+bx_1x_2+x_1x_2y_1y_2. \nn
\eea  %
If we define $\hat\rho_{x_1y_1}$ as a solution of $h_i(\hat x_1,\kappa y_2,\hat y_1, x_2/\kappa)=h_i(x_1,x_2,y_1,y_2)$, and proceed as before, the composition $\varphi=\iota_{12}\circ \hat\rho_{x_1y_1}$ gives us
$$
\hat\varphi:(x_1,x_2,y_1,y_2)\mapsto \left(\frac{(a-x_2y_2)y_2}{b-x_2y_2},y_1+\frac{(a-b)x_2}{a-x_2y_2}, \frac{(b-x_2y_2) x_2}{a-x_2y_2}, x_1-\frac{(a-b)y_2}{b-x_2y_2}\right).
$$
Proceeding as before, we find the map $\hat\varphi$ restricts to the coordinates $u_1, v_1$:
\bea  %
\hat u_1 &=& k-\frac{ku_1}{u_1+v_1}-\frac{(a+u_1)v_1}{b+u_1}\nn\\
&&\qquad\qquad +(a-b)\left(\frac{b(b+k)-u_1^2}{(b+u_1)^2}-\frac{b}{b+u_1}+
\frac{bk^2u_1}{(b+u_1)^2((b+u_1)(u_1+v_1)-ku_1)}\right),\nn\\
   \hat v_1 &=& \frac{u_1(k-u_1-v_1)}{u_1+v_1},  \nn
\eea  %
with the same {\em invariant function} $h_2$ and {\em invariant symplectic form} $\omega$.

On this $2-$dimensional space, the maps $\varphi$ and $\hat\varphi$ commute.

Still using the vector field ${\bf X} = (x_1,-x_2,-y_1,y_2)$, we may consider a different involution $\bar\rho_{x_1y_2}$ as a solution of $h_k(\bar x_1,x_2,y_1,\bar y_2)=h_k(x_1,x_2,y_1,y_2)$ and proceed as before.  We obtain
$$
\bar\varphi:(x_1,x_2,y_1,y_2)\mapsto \left(x_2,y_2+\frac{b}{y_1}-\frac{a}{x_2},x_1+\frac{a}{x_2}-\frac{b}{y_1},y_1\right),
$$
which reduces to
$$
\bar u_1 = -a -u_1+\frac{b u_1}{v_1}+\frac{ku_1}{u_1+v_1},\quad \bar v_1 = a +u_1-\frac{b u_1}{v_1},
$$
which preserves the same function $h_2$ and symplectic form $\omega$ and commutes with {\em both} $\varphi$ and $\hat \varphi$ (in the $u_1,v_1$ plane).  Furthermore, $\hat \varphi$ and $\bar \varphi$ even commute in the $4-$dimensional $x-y$ space.
}\eex

\section{The $6$ Dimensional Case}\label{6d}

We carry out the same procedure in $6$ dimensions to reduce to a map on a $4$ dimensional space with an invariant volume.  To reduce further, we need a second symmetry vector ${\bf X}_1$, which should transform correctly under the map (${\bf X}_1\mapsto \pm {\bf X}_1$) and have zero divergence.

The algebra of symmetries of a given set of functions is easy to find.
Let ${\bf \sigma}=(s_1,\dots , s_{6})$ and solve the $3$ equations ${\bf \sigma}\cdot \nabla h_i=0,\; i=1, \dots, 3$ for $s_4, s_5, s_6$.  Since the functions are independent, this is always possible for some subset of $s_i$ containing $3$ elements.  Generically, this can be the first $3$.  Otherwise, it is always possible to relabel coordinates, so that it {\em is} this case.  This gives us $3$ symmetry vectors of the form
\be\label{syms}  %
{\bf \sigma}_1=(1,0,0,s_4,s_5, s_6),\quad {\bf \sigma}_2=(0,1,0,s_4,s_5, s_6), \quad  {\bf \sigma}_3=(0,0,1,s_4,s_5, s_6),
\ee  %
where, for vector ${\bf \sigma}_i$, $s_{3+k}$ should be replaced by the coefficient of $s_i$ in the solution.  These symmetries commute.  Any linear combination of ${\bf \sigma}_i$ is a symmetry.  The given symmetry vector $\bf X$ is a simple linear combination of these.

\bex[A $6d$ McMillan Map]  \label{mcm6d}  {\em  %
Here we start with ${\bf X} = (x_1,-x_2,x_3,-y_1,y_2,-y_3)$, with invariants
$$
\gamma_1=x_1y_1,\;\; \gamma_2=x_2y_2,\;\; \gamma_3=x_3y_3,\;\; \gamma_4=x_1x_2,\;\; \gamma_5=y_1y_2,\;\; \gamma_6=x_2x_3,\;\; \gamma_7=y_2y_3,
$$
satisfying $\gamma_1\gamma_2=\gamma_4\gamma_5$ and $\gamma_2\gamma_3=\gamma_6\gamma_7$.  We generalise Example \ref{mcm4d} by choosing
\bea  %
&& h_1 = (1-\gamma_4)(1-\gamma_5)-2a\gamma_1=(1-x_1 x_2)(1-y_1y_2)-2 a x_1 y_1, \nn\\
&& h_2 = (1-\gamma_6)(1-\gamma_7)-2a\gamma_2=(1-x_2 x_3)(1-y_2y_3)-2 a x_2 y_2, \nn\\
&& h_3=\gamma_2-\gamma_3=x_2y_2-x_3y_3,\nn
\eea  %
which are evidently invariant under the involution $\iota_{xy}: (x_i,y_i)\mapsto (y_i,x_i), \quad i=1,\dots , 3$.  The equations
$$
h_k(\tilde {\bf x},{\bf y})=h_k({\bf x},{\bf y}),\quad k=1,\dots , 3,
$$
can be solved for $\tilde x_i,\; i=1,\dots , 3$ to obtain the involution $\rho_{x}$ and this can be composed with $\iota_{xy}$ to give
$$
\varphi=\iota_{xy}\circ\rho_{x}:({\bf x},{\bf y})\mapsto  \left(y_1,y_2,y_3,\tilde y_1,\tilde y_2,\tilde y_3,\right),
$$
where
$$
\tilde y_1=
\frac{x_1y_2(-2ay_1+x_2(1-y_1y_2))(1-y_2y_3)}{2ay_2(y_1-y_3)-x_3y_3(1-y_1y_2)(1-y_2y_3)},\quad
 \tilde y_2=-\frac{x_3y_3}{y_2}-\frac{2ay_3}{1-y_2y_3},\quad \tilde y_3 = -\frac{x_2y_2}{y_3}-\frac{2ay_2}{1-y_2y_3},
$$
which is a coupled McMillan map. The Jacobian of this map can be written as $\tilde y_1/x_1$, so that $x_1y_1$ is an invariant volume. Define the volume form $\Omega_6$ by
$$
\Omega_6 = \frac{dx_1\wedge \dots \wedge dy_3}{x_1y_1}.
$$
For this example, we define ${\bf S}_1=x_1{\bf \sigma}_1$, which commutes with $\bf X$, and note
$$
{\bf S}_1  = \left(x_1,0,0,-\, \frac{x_1(x_2+2ay_1-x_2y_1y_2)}{y_2+2ax_1-x_1x_2y_2},0,0\right)\quad\Rightarrow\quad
   \varphi_* ({\bf S}_1)=
    -\, \left(\frac{\tilde y_1 (\tilde y_2+2a\tilde x_1- \tilde x_1\tilde x_2\tilde y_2}{y_1(y_2+2ax_1- x_1x_2y_2)}\right)\, \tilde {\bf S}_1,
$$
so
$$
{\bf X}_1=(x_1y_1(y_2+2ax_1- x_1x_2y_2),0,0,-x_1y_1(x_2+2ay_1-x_2y_1y_2),0,0)
$$
satisfies $\left. \varphi_* ({\bf X}_1)\right|_{(\tilde x,\tilde y)}= - \tilde {\bf X}_1$.  Note that this does not destroy the commutativity, since we have multiplied by an invariant function of the vector field $\bf X$.  Both $\bf X$ and ${\bf X}_1$ satisfy the divergence condition (\ref{divX}).

At this stage we change to coordinates $(u_1,u_2,v_1,v_2,u_3,v_3)=(\gamma_4,\gamma_6,\gamma_1,\gamma_2,h_3,y_3)$,
with Jacobian $-x_1x_2^2x_3y_3=-u_1u_2v_3$, so
$$
\Omega_6= - \frac{du_1\wedge du_2\wedge dv_1\wedge dv_2 \wedge du_3 \wedge dv_3}{u_1v_1u_2v_3}.
$$
We also have ${\bf X}=-v_3\pa_{v_3}$ and ${\bf X}_1=v_1(v_2+2au_1-u_1v_2)\pa_{u_1}+\frac{v_1(v_1v_2-u_1^2)}{u_1}\, \pa_{v_1}$, so
$$
\Omega_5={\bf X}\;\lrcorner\; \Omega_6=-\, \frac{du_1\wedge du_2\wedge dv_1\wedge dv_2 \wedge du_3}{u_1v_1u_2},
$$
which is evidently closed.  We have $\Omega_4={\bf X}_1\;\lrcorner\; \Omega_5$, is given by
$$
\Omega_4=\left(-\left(\frac{v_1v_2-u_1^2}{u_1^2u_2}\right)\, du_1+\left(\frac{v_2+2au_1-u_1v_2}{u_1u_2}\right)\, dv_1\right)\wedge du_2\wedge dv_2\wedge du_3= -dr\wedge du_2\wedge dv_2\wedge du_3,
$$
where
$$
r=\frac{u_1v_1v_2-u_1^2-2au_1v_1-v_1v_2}{u_1u_2}.
$$
In these coordinates
$$
h_1=1-u_1-2av_1+v_1v_2-\frac{v_1v_2}{u_1}=u_2r+1, \quad h_2=1-u_2-2av_2-\frac{v_2(u_3-v_2)(u_2-1)}{u_2}.
$$
Writing $u_2=(h_1-1)/r,\; v_2=s$, we have
$$
 \Omega_4 = \frac{dr\wedge ds \wedge dh_1\wedge du_3}{r}
$$
and (with $h_1=k_1,\, h_3=k_3$)
$$
h_2=\frac{(k_1-1)(1-k_1+r)-rs((k_1-1)(2a+k_3)-k_3r+s(1-k_1+r))}{(k_1-1)r}.
$$
In the $r-s$ plane, the map reduces to
$$
\tilde r = \frac{(k_1-1)^2}{rs(s-k_3)},\quad \tilde s = (k_3-s)\left(1+\frac{2ars}{k_1-1+(k_3-s)rs}\right),
$$
which is symplectic with respect to
$$
\omega= \frac{dr\wedge ds}{r}
$$
and has $h_2$ as an invariant.  This map is exactly the QRT map built out of $h_2$.

We now choose ${\bf X} = (x_1,-x_2,-x_3,-y_1,y_2,y_3)$, with invariants
$$
\gamma_1=x_1y_1,\;\; \gamma_2=x_2y_2,\;\; \gamma_3=x_3y_3,\;\;
\gamma_4=x_1x_2,\;\; \gamma_5=y_1y_2,\;\; \gamma_6=x_2y_3,\;\; \gamma_7=x_3y_2.
$$
Proceeding as before leads to $\hat \varphi=(y_1,y_2,y_3,\hat y_1,\hat y_2,\hat y_3)$, where
\bea  %
&& \hat y_1 =
\frac{x_1y_2^2y_3(x_2(1-y_1y_2)+2ay_1)}{y_2^2+2ay_2y_3+y_3^2-y_2(y_1y_2^2+x_3y_3^2+y_1y_3^2)+x_3y_1y_2^2y_3^2},\nn\\
&& \hat y_2=\frac{1}{y_3}+\frac{y_3(1-x_3y_2)}{y_2^2}+\frac{2a}{y_2},\quad
\tilde y_3 = \frac{1}{y_2}+\frac{y_2(1-x_2y_3)}{y_3^2}+\frac{2a}{y_3},  \nn
\eea  %
which preserves the same volume as above.  The same vector field ${\bf X}_1$ is used and leads to the $2$ dimensional reduction
$$
\hat r = \frac{(k_1-1)^3}{r^2s^2+(k_1-1)(k_1-1+rs(2a-s))},\quad
\hat s = \frac{(k_1-1)^3+k_3(k_1-1)rs\hat r}{(k_1-1)\hat r rs},
$$
where we have shortened the formula for $\hat s$ by using the definition of $\hat r$.  This map is symplectic with respect to the same
$\omega$ and preserves the same function $h_2$ and commutes with the QRT map $\varphi$.
}\eex  %

\section{Summary and Conclusions}\label{summary}

In this paper we have considered a class of mapping on a $2n-$dimensional space (for the cases $n=2,3$), with coordinates $({\bf x},{\bf y})=(x_1,\dots ,y_n)$, possessing $n$ independent functions $h_k({\bf x},{\bf y}),\; k=1,\dots ,n$, which are invariant under the action of the map.   The functions $h_k$ were built out of the invariants of a simple vector field $\bf X$, so are naturally defined on the $(n-1)-$dimensional space of invariants. When the map has invariant (or anti-invariant) volume $\Omega$, with respect to which $\bf X$ is divergence free and when ${\bf X}\mapsto \pm {\bf X}$ under the map, then ${\bf X}\;\lrcorner\;\Omega$ is an invariant volume on this $(n-1)-$dimensional space of invariants.
Since the vector ${\bf X}\mapsto \pm {\bf X}$, the map splits into an $(n-1)-$dimensional map on its space of invariants, together with a map of $v_n$, with coefficients depending on the other variables.

The case $n=2$ is straightforward, with the general structure described in Section \ref{general4d}.  Looking at Example \ref{mcm6d} we see that when $n=3$ we have first to find a second symmetry vector with good transformation and divergence properties.  If we find this, then, as in Example \ref{mcm6d}, we can reduce to a $4-$dimensional space whose volume element has special structure
$$
{\bf X}_1\;\lrcorner\;{\bf X}\;\lrcorner\;\Omega_6=\Omega_4=\omega\wedge dh_{i_1}\wedge dh_{i_2},
$$
where $\omega$ is an invariant $2-$form and $h_{i_k}$ are two of the invariant functions, taken as coordinates.  Since $\Omega_4$ is non-degenerate, $\omega=dr\wedge ds/\sigma(r,s)$, with $r, s, h_{i_1}, h_{i_2}$ being independent coordinates, and the mapping will reduce to the $r,s$ space, preserving $\omega$ and the remaining invariant function.  The third independent symmetry vector is then the Hamiltonian vector field generated by this invariant function.  An important part of our construction is the use of {\em different} starting vectors ${\bf X}$ to produce commuting maps in the $2-$dimensional space.  Commuting maps are an important feature of integrability (see Veselov's review \cite{91-4}).

We have given a systematic {\em procedure} for analysing the maps for a given set of functions $h_k$, built from the invariants of a vector field $\bf X$.  However, there are a number of potential obstacles en route. First, for a given set of functions, there are many possible involutions to consider and some of these are not in rational form.  Some don't have an obvious invariant volume form.  When $n\geq 3$ we need to construct additional symmetry vectors which have the correct transformation and divergence properties.  There is no guarantee that such vectors exist, since we can't expect all $2n-$dimensional maps to reduce to $2-$dimensions, and even when the vectors exist, they may be difficult to find.  We do not know why the maps we construct should commute, but it is interesting that they do.  It may be possible to construct further commuting maps, as we did in Examples \ref{mcm4d-2} and \ref{yb2-ex}, but we have no systematic way of working through the list and no way of classifying all possible reduced maps for a given collection of functions $h_k$.

Most of our examples were {\em constructed} in order to illustrate the technique.  However, Examples \ref{ay-ex} and \ref{yb2-ex} show that these are not just artifacts of our construction, but actually arise in other contexts, such as in the study of Yang-Baxter maps.  The fact that these reduce to $2-$dimensions and possess commuting maps is interesting and previously unknown.

\subsubsection*{Acknowledgments:}
The authors would like to thank the Centro Internacional de Ciencias in Cuernavaca for hospitality in November and December of 2012, during which much of this work was done.  PK was supported by the Australian Research Council Discovery Grant No. DP110102001.


\end{document}